\documentclass[floatfix, preprintnumbers, amsmath, amssymb, 10pt]{revtex4}
\usepackage{dcolumn}
\usepackage{bm}
\begin{document}

\newcommand{\eq}[2]{\begin{equation}\label{#1}{#2}\end{equation}}
\date{\today}
\preprint{CPHT-RR 138.1209}
\preprint{LPTENS-10/13}
\title{On Semi-classical Degravitation and the Cosmological Constant Problems}
\author{Subodh P. Patil$^{1,2)}$}

\affiliation{1) Laboratoire de Physique Th\'eorique\\ Ecole Normale Sup\'erieure,\\ 24 Rue Lhomond, Paris 75005, France\\}
\affiliation{2) Centre de Physique Th\'eorique\\Ecole Polytechnique and CNRS,\\Palaiseau cedex 91128, France}
\email{patil@cpht.polytechnique.fr}

\begin{abstract}
In this report, we discuss a candidate mechanism through which one might address the various cosmological constant problems. We first observe that the renormalization of gravitational couplings (induced by integrating out various matter fields) manifests non-local modifications to Einstein's equations as quantum corrected equations of motion. That is, at the loop level, matter sources curvature through a gravitational coupling that is a non-local function of the covariant d'Alembertian. If the functional form of the resulting Newton's `constant' is such that it annihilates very long wavelength sources, but reduces to $1/M^2_{pl}$ ($M_{pl}$ being the 4d Planck mass) for all sources with cosmologically observable wavelengths, we would have a complimentary realization of the degravitation paradigm-- a realization through which its non-linear completion and the corresponding modified Bianchi identities are readily understood. We proceed to consider various theories whose coupling to gravity may a priori induce non-trivial renormalizations of Newton's constant in the IR, and arrive at a class of non-local effective actions which yield a suitably degravitating filter function for Newton's constant upon subsequently being integrated out. We motivate this class of non-local theories through several considerations, discuss open issues, future directions, the inevitable question of scheme dependence in semi-classical gravitational calculations and comment on connections with other meditations in the literature on relaxing of the cosmological constant semi-classically.
\end{abstract}

\maketitle

\section{Preliminary Remarks}

As is widely appreciated, the cosmological constant (CC) problem \cite{ccp} is actually three devilishly difficult problems rolled into one. To wit, one has to explain why its natural value is not of order unity in Planck units (the so called bare cosmological constant problem) as well as explaining why it is not zero (to one part in $10^{-120}$) as current observations suggest \cite{ccrev}, in addition to explaining why it has only started to dominate the energy budget of the universe recently (the coincidence problem).

Perhaps the simplest manner in which to appreciate the logical independence of the three cosmological constant problems is think of them in terms of their consequences for the lifetime and the structure of the universe. The bare cosmological constant problem, posed differently, is simply asking the question why the universe evolves at all rather than staying deep within the quantum gravity regime, composed of causally disconnected regions with horizons comparable to the Planck size. To ask why it is not zero, is to ask why the universe asymptotes to a de Sitter phase in the future. The coincidence problem is to ask why the observed transition from matter dominated evolution to the de Sitter phase happens when it does, so that several cycles of stellar creation and destruction have taken place (a process that takes of the order of billions of years), so that the heavier elements and the simple molecules that are a pre-requisite for the evolution of complex organisms have been synthesized \footnote{Although the last two aspects of the CC problem might seem related, they are in fact independent as the coincidence problem reflects the fact that one has to tune the matter density relative to the vacuum energy density to one part in $10^{35}$ at the end of inflation in order to attain the ratio $\rho_\Lambda/\rho_m \sim 1$ in the present epoch.}.

At present, two broad categories of approaches towards resolving the cosmological constant problem can be distinguished. On the one hand there is the anthropic approach (see \cite{anth} for a partial survey), and on the other, a varied set of musings in the literature on the possible dynamical resolution of the cosmological constant problem (see \cite{dynamical}\cite{poly} for a review). In particular, it seems reasonable to assume that any attempt that does not resort to anthropic reasoning would {\it have} be dynamical in nature, especially in considering the coincidence problem which appears to demand some form of tracking with matter or some form of relaxation mechanism to alleviate it. We do not wish to delve into the relative merits of either approach in this report, whose focus is simply to present a candidate solution to the CC problems that arises in semi-classical gravity. This approach-- semi-classical degravitation-- also qualifies as a dynamical approach to solving the CC problems, and the realization that we wish to present in this paper will turn out to have overlaps with several other approaches \footnote{There is also the attitude that discounts the bare cosmological constant problem completely, but attempts to explain dark energy and its matter tracking (thus resolving the co-incidence problem) via quitessence, and related models \cite{quin}}. We note in passing that in principle, it would be possible to immediately distinguish between any putative anthropic or dynamical resolution to the CC problems through any signatures of a time dependent component to dark energy. Such an observation would strongly disfavour anthropic scenarios (see \cite{dde} for interesting indications in this regard).  

We commence this report by reviewing a mechanism which has been known for some time, but whose exact realization and exact interpretation when taken as a non-local modification to Einstein's equations remain open questions-- degravitation \cite{dgs}\cite{addg}, when taken as a phenomenological paradigm, attempts to solve the bare cosmological constant (CC) problem directly by nullifying the gravitational effects of the bare CC no matter how large its value in Planck units. In doing so it also contains mechanisms to explain what we observe as dark energy in the present epoch as a memory effect, in addition to considerably alleviating the coincidence problem \cite{sp}. The manner it achieves all of this is to posit that Newton's constant is not so much a universal constant so much as it is a scale dependent coupling such that its strength for a source with a characteristic wavelength $\lambda$ satisfies:
\begin{eqnarray}
\label{sd}
M^2_{pl}G_N(\lambda,L) \to &0&~~~  \lambda \gg L,\\
\nonumber M^2_{pl} G_N(\lambda,L) \to &1&~~~  \lambda \ll L.
\end{eqnarray}
where $L$ is some IR length scale (typically taken to be much larger than any cosmologically accessible scale), and $M_{pl}$ is the usual four dimensional Planck mass. Specifically, Newton's constant is to be thought of as a filter function
\eq{gn}{G_N \to G_N(L^2\square)}
that satisfies (\ref{sd}), and acts on sources of energy-momentum. In \cite{dhk} it was shown that linear perturbations away from Minkowski space for massive, or resonantly massive gravitational theories coupled to conserved sources results in the parametrized filter function:
\eq{rg}{8\pi G_N \to \frac{8\pi G_N}{1 + (\frac{m^2}{\square})^{1-\alpha}}, ~~0\leq \alpha < 1,}
where $\alpha = 0$ parametrizes massive gravity and $\alpha = 1/2$ corresponds to the corrections to the brane graviton propagator in a DGP-like setup. The bounds on $\alpha$ in the above follow from unitarity \cite{gdr}. In the case of massive gravity perturbed around Minkowski space, such a filter function arises when we integrate out the extra polarizations of the graviton, such that the spin 2 part of the metric tensor couples to conserved sources with the resulting filter function (\ref{rg}) (to be understood as an expression only perturbatively). This work was subsequently followed up and generalized in \cite{new}\cite{new2} to address some of the open issues of the constructions considered in \cite{dhk}. Further developments seem to imply natural realizations of the degravitation paradigm (though with drastically different functional forms for the resulting filter functions) in higher co-dimensional embeddings \cite{jac}\cite{penn}. 

We also observe that, although designed to nullify the bare CC by default, the non-local nature of any suitably degravitating filter function (\ref{rg})-- provided we can take its non-linear completion and its existence around a wide class of backgrounds seriously-- also contains mechanisms to explain the magnitude of extremely small value of the apparent CC today as well as considerably alleviating the coincidence problem \cite{sp}; a review of these aspects of degravitation is offered in appendix A. As such, degravitation as a paradigm offers a candidate mechanism which has the potential to address aspects of each of the cosmological constant problems through only the simple assumption that the gravitational coupling $G_N$ vanishes for asymptotically long wavelength sources. 

Motivated by this, we investigate in this report the possibility that degravitation might be nothing other than the configuration space manifestation of the running of gravitational couplings (see \cite{hamber1} for an earlier application of this observation). In this way, its existence around arbitrary backgrounds, its non-linear completion and the modified Bianchi identities implied by (\ref{gn}) are readily understood, and are precisely analogous to well understood phenomenon in conventional gauge theories. That is eq. (\ref{rg}), and generalizations thereof, are to be thought of as arising from renormalization group (RG) flow. 

We begin this report by reviewing how running couplings can imply non-local equations of motion as a means to encode semi-classical physics in gauge theories. We review how familiar quantum phenomenon such as vacuum polarization and loop corrections to effective potentials can be viewed as nothing but non-local modifications to the equations of motion satisfied by the sources and fields at hand. We consider the same situation when we bring gravity in to the picture, later motivating ourselves with a suggestive 2-dimensional example. We also discuss how the modified Bianchi identities are to be understood in this context. Furthermore, in considering the modifications to the graviton propagator implied by renormalizations of the form (\ref{rg}), we understand through the spectral representation of the renormalized propagator that degravitation, thus realized, is nothing other than the result of having either generated a mass gap for gravity under RG flow (as would be the case for a filter function of the massive gravity type ($\alpha = 0$)), or the generation of an infinite tower of massive resonances \cite{gdr}\footnote{Notably, we directly encounter here the fact that diffeomorphism invariance (and the resulting gravitational Ward identities) do preclude shifting the zero momentum pole of the graviton propagator, or the generation of extra poles under RG flow in the same way that gauge invariance protects gauge boson masses under RG flow via the gauge Ward identities. That is, we can consistently generate an effective graviton mass, or extra massive resonances (provided the unitarity bounds on $\alpha$ in (\ref{rg}) are satisfied) around non-trivial backgrounds under RG flow fully consistent with diffeomorphism invariance \cite{pc}. See also the footnote at the end of section IIb)}. Having understood the contours of the program we are interested in executing, we then outline our calculational approach to the general problem, paying attention to its potential pitfalls and open issues. 

We then proceed to review how gravitational couplings are renormalized upon integrating out various matter contents using standard heat kernel techniques. Realizing that any field content that induces non-trivial running of $G_N$ deep into the IR would itself have to manifest non-trivial physics in the IR, we then generalize the heat kernel technique to incorporate field theories with non-standard kinetic terms and consider various classes of theories that feature novel IR physics. We first orient ourselves with examples corresponding to DGP inspired setups and non-commutative field theories, among others. Drawing lessons from these examples we are led to a class of non-local field theories that formally exhibit degravitation and nullifies the cosmological constant. We motivate this class of non-local field theories through several contexts in which it could arise. In a particular scheme, we infer degravitation of the massive gravity type (with $\alpha = 0$ in (\ref{rg})), but where now the non-linear completion of the theory and the implied modified Bianchi identities are readily understood. We temper this result with a discussion of the inevitable scheme dependence that exists in any such calculation. Although the existence of a degravitating solution appears to be a scheme independent conclusion, the precise functional form of the resulting filter function appears not to be. We view this as an inevitable consequence of the limitations of our calculational techniques rather than an intrinsic aspect of the general program we outline in this report-- that non-trivial RG flow in the IR results in a non-local filter function as the appropriate gravitational coupling (\ref{gn}). Therefore the perspective of the results contained in this paper is that we present the basis of a novel realization of the degravitation paradigm (through which the cosmological constant problems are naturally addressed), through which its non-linear completion and existence around arbitrary backgrounds are readily understood, and furnish preliminary evidence for this claim.

\section{Filtering gravity by running $G_N$}

In a typical gauge theory, it has been observed that the result of loop corrections in position space, is to effect the replacement for the coupling constants \cite{barvinsky}:

\eq{run}{\alpha \to \alpha_{0}(\mu) + c~ ln [-\square/\mu^2],}
where $\mu$ is some (experimentally determined) reference energy scale (such that $\alpha(\mu) = \alpha_0$), and the coefficient $c$ depends on the renormalization group equations of the theory at hand. In fact, it is relatively straightforward to infer from the basic properties of Fourier transforms, that since the bare coupling constant multiplies a product of local operators in the bare action, the effects of RG running would be to induce some form of non-locality in the vertex at the level of the semi-classical spacetime action once we transform back from momentum space \cite{mazitelli}. That is, the net effect of quantum corrections to a physical system can be encoded by non-local modifications to the low energy effective action, which is then used to extract the correct semi-classical physics. This situation is certainly familiar to the reader, as we illustrate with an example from QED, where the interaction Lagrangian reads:

\eq{lint}{\mathcal L_{int } = e\bar\Psi \gamma^\mu A_\mu \Psi(x).}
In Fourier space, it is clear that if our coupling runs as a function of the scale (set by the external momentum $p$) of the process at hand (i.e. $e \to e(p^2/\mu^2)$), then we can just as easily rewrite the coupling as $e(\square/\mu^2)$ (where the delta function constraint on the vertex results in $\square \to p^2$). This suggests that in position space, we can encode quantum effects at the level of the action through the substitution:
\eq{lintbox}{\mathcal L_{int } = e(\square/\mu^2)\bar\Psi \gamma^\mu A_\mu \Psi(x).}
To flesh this out, consider that in QED, it is straightforward to show that corrections to the photon propagator from virtual fermion loops results in the running \cite{huang}
\eq{ferml}{\alpha \to \alpha(k^2) = \alpha\Bigl[1 + \frac{\alpha}{3\pi}ln~\frac{-k^2}{m_e^2} + O(e^4_0)\Bigr] ~~; k^2/m_e^2 >> 1,}  
where $\alpha$ is the low energy fine structure constant $\alpha \approx 1/137$. This implies the non-local form factor (\ref{run}) in configuration space:
\eq{fermlnl}{\alpha \to \alpha\Bigl[1 + \frac{\alpha}{3\pi}ln~[-\square/m_e^2] + O(e^4_0)\Bigr],}
Indeed, we know that something of the type has to in fact happen at the semi-classical level if we are to obtain the correct corrections to the classical equations of motion from quantum effects. In QED, we know that the electrostatic potential felt by an electron due to the presence of an infinitely heavy point charge (with momentum transfer $k^\mu = (0,\vec k)$) is given by the Fourier transform of the scattering amplitude:

\eq{pot}{V(r) = \int \frac{d^3k}{(2\pi)^3} e^{i\vec k\cdot \vec r}\frac{4\pi\alpha(- k^2)}{k^2}.}
If we took the fine structure were truly constant ($4\pi \alpha = e_0^2$), then the above evaluates to the usual classical answer:

\eq{classans}{V(r) = \frac{e_0^2}{4\pi r}.}
However once we account for the fact that couplings do in fact run, we realize that there must be a deviation from Coulomb's law once quantum effects are incorporated. It can be shown \cite{huang}, that such a deviation can be accounted for through the fact that quantum mechanically, the electron effectively gets smeared out into the charge distribution:

\eq{ecd}{\rho(r) = \int \frac{d^3k}{(2\pi)^3} e^{i\vec k\cdot \vec r}4\pi\alpha(- k^2),}
which is accounted for in (\ref{pot}) through the prescription (\ref{run}):

\eq{nlpot}{V(r) = \int \frac{d^3k}{(2\pi)^3} 4\pi\alpha(- \square) e^{i\vec k\cdot \vec r}\frac{1}{k^2},}   
so that it is straightforward to compute the net effect of the running of $\alpha$ (\ref{ferml}) to be to give us the quantum corrected potential in the limit of a vanishing electron mass (i.e. for $mr \ll 1$):

\eq{qcpot}{V(r) =  \frac{e_0^2}{4\pi r}\Bigl[1 + \frac{\alpha}{3\pi}\Bigl( ln\frac{1}{m^2r^2} -2\gamma \Bigr)\Bigr],}
where $\gamma$ is the Euler constant ($-2\gamma \approx -1.154$). Through this example, we specifically see how quantum effects alter the classical equations of motion, such that the original QED interaction Lagrangian has to be altered in the manner suggested by (\ref{run}) to capture semi-classically, the effects of vacuum polarization. Specifically, one could either view this as the Gauss law (which in the classical equations of motion, results in the potential (\ref{classans}) for a point charge) being modified to reflect (\ref{qcpot}) as the correct potential experienced by an electron, or equivalently the smearing out of the electron into a charge distribution given by (\ref{ecd}). Either perspective follows from (\ref{run}). In a similar fashion, we are interested in computing semi-classical corrections to the gravitational action, such that the degravitation paradigm might find a realization as a semi-classical effect with (\ref{gn}) being the resulting transcription of (\ref{run}). In doing so, we find that we find a natural accounting for the modified Bianchi identities implied by promoting Newton's constant to a filter function-- an issue which presently deserves a quick digression. 

Considering the modified Einstein equations implied by (\ref{gn}) \cite{addg}
\eq{mee}{G^\mu_\nu = 8\pi G_N(L^2\square)T^\mu_\nu,}
one immediately encounters the issue, that since in general $[\nabla_\mu,\square] \neq 0$, either the Bianchi identities are not satisfied in the above, or new constraints are imposed on a covariantly conserved energy momentum tensor (recalling that any covariantly constructed field theory will yield a covariantly conserved energy momentum tensor). Although this might cause some readers anxiety, our perspective here is that the correct Bianchi identity is in fact $\nabla_\mu [G(L^2\square)T^\mu_\nu] = 0$, and that this is to be expected considering the fact that in reality, the equations (\ref{mee}) are actually a paraphrasing of the semi-classical equations of motion:
\eq{semi}{G^\mu_\nu = \frac{1}{M^2_{pl}}\langle T^\mu_\nu \rangle.}
That is, one obtains the energy momentum tensor by taking the variational derivative with respect to the metric tensor {\it of the effective action} (rather than the classical action):
\eq{effact}{\langle T_{\mu\nu} \rangle = \frac{2}{\sqrt{-g}}\frac{\delta W}{\delta g^{\mu\nu}}.}
Where in the example of having integrated out a scalar degree of freedom \cite{bd}:
\begin{eqnarray}
\label{effld} W &=& \int d^4x \sqrt{-g(x)}~\mathcal L_{eff}(x)\\ 
 \mathcal L_{eff} &=& \frac{i}{2}\lim_{x'\to x}\int^\infty_{m^2}d\bar m^2 G_F^{DS}(x,x';\bar m^2),
\end{eqnarray}
with $G^F_{DS}(x,x';\bar m^2)$ being the Schwinger-De Witt representation of the Feynman propagator of a scalar field with mass $\bar m$ \cite{bd}. As we have seen in our previous example, just as loop corrections induce non-local operators (\ref{run}) in place of coupling constants, which effects the intuitive idea that quantum effects `smear' out classical sources of fields (\ref{ecd}) or provide quantum corrections to the classical equations of motion, so the running of Newton's constant will result in (\ref{mee}) as a paraphrasing of (\ref{semi}) \footnote{We note that non-local contributions to $\mathcal L_{eff}$ from loop corrections can have several sources, including quantum corrections to the quantum state defining the expectation value (\ref{effact}), in addition to contributions which remain finite in the limit $x'\to x$ in (\ref{effld}) (i.e. are sensitive to the large scale structure of the manifold)\cite{bd}.}. Therefore, it should be clear that the quantum corrected equations of motion derived from the effective action ($\mathcal L_{eff}$) are implied by $\nabla_\mu [G(L^2\square)T^\mu_\nu] = 0$, as $\nabla_\mu T^\mu_\nu = 0$ implies only the classical equations of motion, derived from the classical Lagrangian density, $\mathcal L_{c}$.

\subsection{An example in $2+ \epsilon$ dimensions}

In pursuit of further motivation, we discuss a toy example \cite{hamber} which arises in the context of 2-d gravity-- where gravity is a renormalizable theory. Duff and Christensen \cite{duff} have computed the one loop beta function when gravity is coupled to $n_s$ massless spin s particles, and $N_s$ massive spin s particles in $D = 2+\epsilon$ dimensions as:

\eq{bgrav}{\beta(G) = (d-2)G - \beta_0G^2 - ...,}
where $\beta_0$ is given by
\eq{b0}{\beta_0 = \frac{2}{3}[1 - n_{3/2} + n_{1/2} - n_0 - N_1 + N_{1/2} - N_0],}
From (\ref{bgrav}) we can immediately read off the existence of a UV fixed point for gravity at $G_c = \epsilon/\beta_0$, if $\beta_0>0$. The differential equation defining the beta function:
\eq{diffeq}{\mu\frac{dG(\mu)}{d\mu} = \beta(G)}
is readily solved to yield the scale dependence of $G_N$ as:
\eq{runga}{G(k^2) = \frac{G_c}{1 \pm (m^2/k^2)^{\frac{\epsilon}{2}}},} 
where $m^2$ is constant of integration, determined at the energy scale where we experimentally measurre $G_N$ through some local experiment. In configuration space, this would result in the promotion of Newton's constant to the filter function:
\eq{ff}{G(\square) = \frac{G_c}{1 \pm (m^2/\square)^{\frac{\epsilon}{2}}},}
which suggestively compares with the  family of filter functions proposed in \cite{dhk} (\ref{rg}):
\eq{mgff}{G(\square) = \frac{G_c}{1 + (m^2/\square)^{1-\alpha}}.}
In what follows, we set about the endeavour of obtaining an appropriately degravitating filter function for gravity in 4-dimensions. Before we do so, we digress momentarily on the physical interpretation of a running Newton's constant. 

\subsection{Generating mass gaps and poles under RG flow}

As discussed in \cite{gdr} and emphasized in the context of degravitation in \cite{dhk}, any theory that induces an effective momentum dependence for $G_N$ necessarily looks like a theory that propagates a massive graviton, or many massive resonances. This is most easily seen around a Minkowski background in the context of a filter function of the form (\ref{rg}), which would result in the scalar part of the graviton propagator assuming the form
\eq{pform}{\Delta(k^2) = \frac{1}{k^2 + M^2(k^2)},} 
where the precise functional form of $M^2(k^2)$ depends on the value $\alpha$. We can immediately perform a spectral decomposition of the propagator
\eq{specd}{\Delta(k^2) = \int^\infty_0 dm^2\frac{\rho(m^2)}{k^2 + m^2},}
where, for example, DGP-like propagators (with $\alpha = 1/2$ in (\ref{rg})) admits the spectral representation \cite{dhk}
\eq{dgpspect}{\frac{1}{-\square +M\sqrt{-\square}} = \frac{2M}{\pi}\int^\infty_0 \frac{dm^2}{m^2 + M^2}\frac{1}{m^2 - \square};~~~\rho(m^2) = \frac{2M}{\pi(M^2 + m^2)},}
and the propagator for massive gravity admits the simple spectral representation:
\eq{mgspect}{\frac{1}{-\square + M^2} = \int^\infty_0 dm^2\frac{\delta(m^2 - M^2)}{-\square + m^2}.}
In the case corresponding to a DGP propagator, we see that the spectral decomposition implies that the theory propagates an continuum of massive resonances. For the case corresponding to massive gravity, there is only one isolated pole at $M^2$, as expected. Evidently any momentum dependence of $G_N$ that admits a spectral representation (\ref{specd}) forces us to interpret the resulting theory as propagating massive resonances.

Therefore if the effect of RG flow in the IR is to induce an effective propagator of the form (\ref{pform}) generalized to arbitrary backgrounds, evidently one of two phenomena are occurring. Either the pole at zero momenta remains fixed and extra poles have been generated by renormalization, or in the case of a resulting filter function of the massive gravity type, the pole at $p^2 = 0$ has been lifted and an effective mass gap has been generated in the graviton spectrum. In the former case, any extra isolated poles would have to correspond to new spin 0 states unless there were an infinite tower of them. This is due to the fact that any propagator with a finite number of spin two poles would correspond to a theory with a finite number of higher order curvature invariants which generically propagate ghosts and correspond to unphysical theories. Any physical trajectories in theory space induced by RG flow would have to flow along trajectories that do not admit propagating low energy ghosts. In particular, were they to admit a parametrization for the renormalized Newton's constant that corresponds to (\ref{rg}), they would have to satisfy the unitarity bounds $0\leq \alpha <1$ \footnote{\label{foot} Furthermore, we take note here that it is entirely consistent to generate massive spin two poles under RG flow consistent with diffeomorphism invariance. This is rather unlike the situation in gauge theories where gauge invariance (via the Ward identities) protects gauge boson masses under renormalization. The reason for this is that Einstein-Hilbert gravity is a non-renormalizable theory, and corresponds only to the two first terms in a low energy effective theory expansion which samples a theory space containing many (in principle infinite) higher dimensional operators under RG flow, unlike in renormalizable theories where the theory space consists of a small number of dimensions. Therefore, just as one can write down higher order curvature invariants where the graviton propagator can acquire a spin two pole consistent with diffeomorphism invariance, so a given RG trajectory flow through regions of theory space where extra poles for the graviton propagator result.}. However, the gain in generating (\ref{rg}) through renormalization rather than from any background specific construction, is that its non-linear completion and consistency is guaranteed provided we begin our flow from a consistent (i.e. ghost free) point in theory space. Thus the approach that we espouse in this report considers Einstein Hilbert gravity coupled to various matter fields (that describe consistent theories which manifest novel IR physics), and attempts to study the resulting RG flow of the coupled system in the spirit of pursuing a proof by construction. In this way, degravitation would find a realization as an effect in semi-classical gravity. One particular example that we find results in degravitation is obtained by inferring the RG evolution of gravity coupled to a toy scalar field with the non-local action:
\eq{ssfact}{S = \int d^4x\sqrt{-g}\phi\Bigl[e^{m^2/\square}\square\Bigr]\phi,}
which in spite of its unusual appearances, arises naturally in several contexts and defines consistent propagation of massless quanta with non-trivial boundary conditions as we discuss further. Before we commence with the thrust of our investigation, we take a moment to outline the strategy we will pursue.

\section{The Outline}
The basic question we are attempting to answer in this report, is whether or not gravitational couplings can be engineered to run deep in the IR such that the combination $G_N\Lambda$ flows to zero, and that Newton's constant assumes the form (\ref{gn}) in its action on all other sources of energy momentum. The discussion until now has demonstrated the utility and the desirable phenomenology of such a possibility, and the remainder of this report focusses on attempting to realize this possibility in a proof by construction. 

Clearly we are interested in computing the beta functions for the gravitational couplings in the Einstein Hilbert action, and inferring the renormalization group flow that results. There are several independent ways one can attempt this calculation, and in this report we focus on the particular approach offered by heat kernel techniques. This is not the only way to proceed, and any conclusions drawn in this investigation would ultimately have to be cross checked with other available methods. The latter point assumes further importance when we realize that computing the running of couplings in a non-renormalizable theory (with dimensionful couplings at the lowest order) is bound to involve an irreducible degree of scheme dependence. Although the fact that degravitation is indeed effected in the example we consider will turn out to be independent of the scheme, the precise functional form of the resulting filter function inherits this scheme dependence. 

Therefore we approach this study in the spirit of delineating the idea that degravitation has a rather natural realization as a loop effect in semi-classical gravity, and with a particular technique, coupled to a particular field content, furnish evidence to this effect. Any firm claims in the affirmative, would certainly have to be examined with other techniques (lattice simulations, functional renormalization group techniques etc). With these caveats in mind, we begin our investigation with a review of the heat kernel renormalization technique in curved spacetimes.  

\section{Effective action for Gravity coupled to a scalar field}

In this section, we review the derivation of the effective action for gravity when coupled to a minimally coupled scalar field employing heat kernel techniques. Starting with the action:
\eq{fsfacte}{S = \int d^4x~\sqrt{-g}\Bigl[-\frac{1}{16\pi G}R + \frac{1}{2}g^{\mu\nu}\partial_\mu\phi\partial_\nu\phi \Bigr],}
one can subsequently compute the following loop corrections to the graviton propagator:
\eq{gplc}{\frac{iG_{R}}{k^2} = \frac{iG_B}{k^2} + \frac{iG_B}{k^2}[i\Sigma]\frac{iG_B}{k^2} + ...} 
where $G_B$ is the bare coupling multiplying the Einstein Hilbert action, and $\Sigma$ is given by evaluating the appropriate loop integrals, and (ignoring tensor structure) is of the form
\eq{1pi}{\Sigma = \frac{c}{16\pi^2}k^2\Lambda^2,}
where $c$ is a number of order unity, and $\Lambda$ is the cut-off in our loop integrals. This allows us to infer the renormalized Newton's constant as:
\eq{reng}{\frac{1}{G_R} = \frac{1}{G_B} + \frac{c}{16\pi^2}\Lambda^2.}
The net effect of explicitly integrating out canonical matter fields using heat kernel techniques \cite{reeb}\cite{wl} as we shall see shortly, is to yield the Wilsonian running:
\eq{wrg}{\frac{1}{G(\mu)} = \frac{1}{G(\mu_{ref})} - \frac{\mu^2}{12\pi}(n_0 + n_{1/2} - 4n_1),}
where $n_0$ is the number of real massless scalars, $n_{1/2}$, the number of Weyl fermions, $n_1$, the number of gauge bosons and $\mu$ the running energy scale. One infers from this, the running Newton's constant:
\eq{rnc}{G(\mu^2) = \frac{G(\mu^2_{ref})}{1 + c~G(\mu^2_{ref}) \mu^2}.} 
Although we see immediately that this form is of little use to us in terms of degravitation, the implications of the above have been explored in the past to see whether or not the scale of quantum gravity can be brought down to the threshold of current accelerator energies by invoking many solely gravitationally coupled species \cite{dgkn}\cite{gd}\cite{redi}. To recap the essentials of the derivation of (\ref{rnc}) from functional methods, we first consider the effective action for a scalar field minimally coupled to gravity:
\eq{easf}{e^{-W} = \int \mathcal D \phi e^{{-\frac{1}{8\pi}}\int\sqrt{g}\phi(-\square + m^2)\phi} ~=~ [det(-\square + m^2)]^{-1/2}.} 
Defining the heat kernel:
\eq{hk}{H(\tau) \equiv Tr~e^{-\tau\Lambda} = \sum_i e^{-\tau\lambda_i},}
with $\lambda_i$ defined as the eigenvalues of $\Lambda = -\square + m^2$. The effective action can then be obtained via the functional identity:
\eq{fid}{W = \frac{1}{2}ln~det\Lambda = \frac{1}{2}\sum_i ln\lambda_i = -\frac{1}{2}\int^\infty_{\epsilon^2}d\tau \frac{H(\tau)}{\tau},}
where we are required to regulate the divergent integral with the cut-off in the lower bound of the integral, with the definition of the heat kernel as:
\eq{hkd}{H(\tau) = \int d^4x~ G(x,x;\tau)= \int d^4x~ \langle x|e^{-\tau(-\square + m^2)}\theta(\tau)|x\rangle,}
where the theta function is necessary as the operator in the exponent has eigenvalues that are unbounded from above in Euclidean space. We thus see that $G(x,x';\tau)$ satisfies the differential equation
\eq{pde}{(\partial_\tau - \square_x + m^2)G(x,x';\tau) = \sqrt g\delta(\tau)\delta^4(x,x'),}
which justifies the nomenclature. In flat space, for a massless scalar field, one has the usual result
\eq{gffp}{G(x,x';\tau) = \frac{1}{16\pi^2\tau^2}e^{-\frac{(x-x')^2}{4\tau}}.}
In curved spacetimes, we have to expand in terms of curvatures to obtain the result (see e.g. \cite{wl}\footnote{In \cite{wl}, renormalization over a manifold with a boundary results in the leading correction of $\tau^{3/2}$ rather than the $\tau^2$ as we have in our case.}):
\eq{cshk}{H(\tau) = \frac{1}{16\pi^2\tau^2}\Bigl[\int dV + \frac{\tau}{6}\int R + O(\tau^{2},R^2) \Bigr].}  
A derivation of this result in a different context will be presented shortly. For our immediate purposes, we note from the above and from (\ref{fid}), that the renormalized Newton's constant is given by:
\eq{rghk}{\frac{1}{G_R} = \frac{1}{G_{ref}} + \frac{1}{12\pi \epsilon^2}.}
One can infer the Wilsonian running of Newton's constant by imposing an IR cut-off \footnote{The IR cut-off in the proper time integral of the heat kernel has the net effect of suppressing the integration of all light momenta. There are several schemes one could have utilized towards this effect, but the hard cut-off remains truest to the Wilsonian prescription, and for our purposes has the added advantage of being the easiest scheme to deal with calculationally.}:
\eq{wrgg1}{W = -\frac{1}{2}\int^{\mu^{-2}}_{\epsilon^2} d\tau \frac{H(\tau)}{\tau},}
and computing the renormalized coupling at a reference energy scale, relative to which we compute the renormalized couplings at any other scale, which we can then write in a manner independent of $\epsilon$ as:  
\eq{wr}{\frac{1}{G(\mu^2)} = \frac{1}{G_{ref}} - c\frac{\mu^2}{12\pi},}
where all the $\epsilon$ dependence has been swept into the (renormalized) quantity $G_{ref}$ which we determine experimentally, and where $c$ is some numerical factor that depends on the number of scalar, fermionic and gauge field degrees of freedom. Similarly, one infers the renormalized cosmological constant:
\eq{rcc}{\Lambda_R = \Lambda_B + (N_b - N_f)\frac{c'}{\epsilon^4}.}

If our goal is to engineer a running for Newton's constant such that the gravitational coupling it implies vanishes in the far IR, we encounter the immediate realization that conventional matter and gauge fields do not do the job. Given that any direct measurement of $G_N$ (say by Cavendish type experiments) sets its scale to be $1/M^2_{pl}$ at laboratory scales, we require that any running induced by matter or gauge field couplings be such that it not only has the right sign, but is also strong enough to render gravity a dud force on extremely large scales. One can easily see through (\ref{wrg}) and (\ref{rnc}) that no massless scalar, fermionic or gauge degree of freedom can induce a suitably degravitating running of Newton's constant. The same can easily be shown to be true of massive degrees of freedom as well. These observations are perhaps painfully obvious, as why should renormalization of these fields-- a process which involves the subtraction of infinities due to divergent loop integrals in these theories-- run deep into the IR? The answer is that it does not, unless the theory itself presents us novel physics (RG flow) in the IR. This novel physics would have to be such that were we to somehow obtain the beta function for this theory coupled to gravity, it would integrate to cause $M^2_{pl}G_N$ to run from $0$ in the far infra red, to $1$ at all cosmologically accessible scales. 

Phrased in this manner, we are simply asking the question of whether or not it is possible to engineer a field content such that gravity has a positive beta function from the far infra red to Hubble scales, and such that $M^2_{pl}G_N \to 0$ and $G_N\Lambda \to 0$ in the far infra red. Although the answer to this question in all generality is beyond our means at the moment, we attempt this investigation in the spirit of pursuing a proof of concept by construction. In attempting to understand the class of theories that degravitate by construction, we also uncover hints of the possible underlying physical mechanisms for this non-trivial RG trajectory in the IR which we will elaborate on further. Before we do so, we observe the general features of a theory that would serve our purposes.

\section{Running $G_N$ in the IR}
In curved spacetimes as we shall see shortly, one can compute the curvature expansion of the heat kernel to find the general form:
\eq{ehren}{H(\tau) = \int d^4x~G(x,x;\tau) = \lambda(\tau) \int dV + g(\tau)\int dV~R + O(R^2).}
In order to obtain the effective action, one integrates the above as:
\eq{intch}{W = -\frac{1}{2}\int^\infty_{\epsilon^2} \frac{H(\tau)}{\tau},}
where the cutoff in the lower limit in the integral (\ref{intch}) is necessary to regulate the integral. Noting that the (bare) Einstein-Hilbert action is of the form:
\eq{beh}{S = -\frac{1}{16\pi G_B}\int dV~ R + \frac{1}{8\pi}\int dV~ \Lambda_B,}
we see that the renormalized gravitational couplings are given by:
\eq{gnomes}{\frac{1}{G_R} = \frac{1}{G_B} + 8\pi\int^\infty_{\epsilon^2}\frac{g(\tau)}{\tau},}
\eq{cnomes}{\Lambda_R = \Lambda_B -4\pi\int^\infty_{\epsilon^2} \frac{\lambda(\tau)}{\tau}.}
To obtain the Wilsonian running of Newton's constant, we compute the expression:
\eq{gnomez}{\frac{1}{G(\mu)} = \frac{1}{G_B} + 8\pi\int^{\mu^{-2}}_{\epsilon^2}\frac{g(\tau)}{\tau},}
and expressing $G_B$ in terms of some reference $G(\mu_{r})$ (and hence absorbing the usual infinite renormalizations required by UV divergent loop integrals) at some reference (UV) scale $\mu_r$ via
\eq{gnomex}{\frac{1}{G_B} = \frac{1}{G(\mu_r)} - 8\pi\int^{\mu_r^{-2}}_{\epsilon^2}\frac{g(\tau)}{\tau},}
we can substitute back into (\ref{gnomez}) to compute the running of Newton's constant as:
\eq{rungnomes}{G_N(\mu) = \frac{G_N(\mu_r)}{1 + 8\pi G_N(\mu_r)\int^{\mu^{-2}}_{\mu_r^{-2}}\frac{g(\tau)}{\tau}}.}
The scale $\mu_{r}$ is necessarily a UV scale ($\mu_r^{-2} < \mu^{-2}$), as we can only {\it directly} measure the strength of $G_N$ via Cavendish type of experiments, and as laboratory scale experiments, are in the far UV as far as cosmological and astrophysical scales are concerned \footnote{The reader might be under the impression that we measure $G_N$ via solar system or larger scale observations, but these necessarily involve masses of distant objects whose mass we {\it infer} from other observations. As the usual gauge theory context, the only way to to experimentally determine the strength of a given interaction at a given scale, and hence fix the renormalization conditions, is to scatter particles with known charges off each other and infer the interaction strength. The gravitational analog of this would be to test the gravitational force between particles with {\it known} masses-- i.e. conduct Cavendish experiments.}. 
We also note here, that in order for degravitation to have meaningfully achieved its main purpose, in actuality we require that the product $G_R(\mu)\Lambda_R(\mu)$ vanish in the IR. That is, given that whatever makes Newton's constant run will also make the cosmological constant run in the IR, in order to degravitate the bare cosmological constant, we require that the product $G_N\Lambda$ (which is what would source curvature) vanish in appropriately dimensionless units in the IR. In this way, the goals of the endeavour of this paper might not seem to be so different from the many approaches to relaxing the cosmological constant via RG running that exist in the literature \cite{dynamicalrun}. However, as we have noted in the introduction and in the appendix, solving the bare CC problem (by construction) is but one aspect of degravitation, which is a more general mechanism. Considering $G_N$ as a filter function acting on all other sources of energy momentum (including CC like contributions), has advantageous features for addressing the other two cosmological constant problems, as well as cosmology in general \cite{sp} due to the intrinsically dynamical nature aspect of this prescription.

The perspective one should have then of the sequel is as follows: we consider the effects of integrating out a particular scalar field in some hidden (dark) sector (i.e. this field couples only minimally couples to gravity and nothing else). We are then in fact, obliged to integrate out this field when considering the low energy interactions of all other fields coupled to gravity. Upon integrating out this field, we will have satisfied our objectives if $G_N \Lambda/M^2_{pl}$ runs to zero, and the resulting $G_N$ becomes a filter function when acting on all other sources of energy density (treated classically). In this way, any other contributions to the cosmological constant are also dynamically rendered gravitationally inert in such a manner that can also address the other two cosmological constant problems (see appendix A). We orient ourselves first with two familiar examples.

\section{Scalar Theories with IR Modified Propagators}

In considering which theories when coupled to gravity might render non-trivial RG flows for the gravitational couplings in the IR, it is a fair assumption to draw upon theories which in and of themselves exhibit, or reflect novel physics in the IR. One such example is furnished by a minimally coupled a scalar field with the Euclidean signature propagator:
\eq{modprop}{\frac{1}{\square - M\sqrt{-\square} - m^2}.}
Such a modified propagator naturally arises on a DGP brane \cite{DGP} as a result of accounting for bulk propagation in the effective theory (with an explicit brane mass term included). We generalize the standard heat kernel technique for computing renormalized couplings induced by such a modified propagator in appendix B. There, we derive that the function $\lambda(\tau)$ which determines the renormalization of the cosmological constant through (\ref{ehren}) is given by (\ref{gfa}):
\eq{gfaxx}{\lambda(\tau) = \frac{e^{-m^2\tau}}{16\pi^2} \Bigl[ \frac{1}{\tau^2} + \frac{M^2}{4\tau} \mp M \frac{e^{\frac{M^2\tau}{4}}\sqrt{\pi\tau}(6 + M^2\tau) [1 \mp Erf (M \sqrt\tau /2)]} {8\tau^2}\Bigr],}
where the signs $\pm$ correspond to the choice of the propagator $(\square \mp M\sqrt{-\square} - m^2)^{-1}$, with the lower sign corresponding to the so-called `self-accelerated branch'. Correspondingly, $g(\tau)$, which determines the renormalization of $G_N$ through (\ref{gnomes}) is given by:
\eq{gtab}{g(\tau) = \frac{e^{-m^2\tau}}{16\pi^2}\Bigr[ \frac{1}{6\tau} + \frac{M^2}{96} \mp \frac{M\sqrt{\pi \tau}}{32\tau}e^{M^2\tau/4}\Bigl(1 \mp Erf[M\sqrt\tau/2]\Bigr)\Bigl(3 + \frac{M^2 \tau}{6}\Bigr)\Bigl].}

We note from (\ref{rungnomes}), that in general, in order to achieve the running $G(\mu)/G(\mu_r) = G(\mu)M^2_{pl}\to 0$ in the IR (as $\mu\to 0$), we require that the integrand $g(\tau)/\tau$ decay no faster than $1/\tau$ at large $\tau$. From (\ref{gtab}), we note that the renormalization of $G_N$ at IR scales vanishes at large $\tau$ for the choice of the propagator $(\square -M\sqrt{-\square})^{-1}$, as can be inferred from the asymptotic formula for large $\tau$ \cite{AS}:
\eq{assss}{1 - Erf[M\sqrt\tau/2] \to \frac{2}{\sqrt{\pi t}M}e^{-M^2\tau/4}\Bigl[1 - \frac{2}{M^2 \tau} + O(\tau^{-2})\Bigr].}
The so-called self-accelerated branch however, formally exhibits the requisite behaviour in the IR. When we tune the brane mass $m$ of our model to be such that $m^2 = M^2/4$ (the choice of which we elaborate upon shortly), the dominant IR (large $\tau$) behaviour of (\ref{gtab}) goes as $M^3/96\cdot\sqrt{\pi \tau}/16\pi^2$. This would result, after obtaining the effective action via (\ref{intch}) and (\ref{beh}) in a filter function for long wavelength sources of the form:  
\eq{finffsr}{{G_N(M^{-2}\square) = \frac{G_N(\mu_r)}{1 + \sqrt\pi G_N(\mu_r)\frac{M^2}{96} \Bigr[\frac{M}{\sqrt{-\square}}  - \frac{M}{\mu_r}\Bigl]}}.}
By defining $\bar G$ as
\eq{finff3sr}{\bar G_N(\mu_r) = \frac{G_N(\mu_r)}{1 - \sqrt{\pi}\frac{M^2}{96}G_N(\mu_r){\frac{M}{\mu_r}}},}
we can rewrite the above as
\eq{finff2sr}{G_N(M^{-2}\square) = \frac{\bar G_N(\mu_r)}{1 + \sqrt{\pi}\frac{M^2}{96}\bar G_N(\mu_r)\frac{M}{\sqrt{-\square}}}.}
We note with interest how the above compares to (\ref{rg}) for the case $\alpha = 1/2$ \cite{dhk}. We can also compute how the cosmological constant runs via (\ref{gfa}) as:
\eq{ccrun}{\Lambda(\mu) = \bar\Lambda(\mu_r) + \frac{M^3}{8\pi}\sqrt{\pi}\mu,}
with
\eq{lbccrun}{\bar\Lambda(\mu_r) = \Lambda(\mu_r) - \frac{M^3}{8\pi}\sqrt{\pi}\mu_r,}
recalling that the infinite renormalizations of the CC coming from the standard UV divergences are all contained in $\Lambda(\mu_r)$. From (\ref{finff2sr}) and (\ref{ccrun}), it is clear that $G(\mu)\Lambda(\mu)/M^2_{pl} \to 0$ in the IR, with the degravitating filter function (\ref{finff2sr}) resulting upon all other sources.

In spite of its appearances however, this example is unphysical. We observe in this that the physical origins of this apparent degravitation lie firmly in a new class of IR divergent loop diagrams. Upon cavalierly computing the one loop effective action given the form of the DGP propagator on the self-accelerated branch, we find the loops responsible for the Wilsonian running that appears to cause the vanishing of Newton's constant in the IR are precisely those diagrams that correspond to poles on the positive real axis in momentum space. In particular, consider one of the diagrams that contributes to the one loop effective action with external graviton legs and internal scalar field momentum $k$:
\eq{oldd}{\Delta\Gamma_1 \sim \int d^4k~\frac{1}{k^2 \pm M k + m^2}}
It is clear that we can write the denominator as $(k-\omega_1)(k-\omega_2)$, with the $\omega_i$ being real and positive roots on the self accelerated branch. The choice $m^2 = M^2/4$ makes the denominator take the form of a perfect square $(k - M/2)^2$, preventing us of regularizing the integral via any sort of principal value prescription. Therefore, we understand this apparent degravitation as arising from IR divergent loops due to states with the dispersion relation $k^2 = M^2/4$. Although these states correspond to plane wave solutions of our field equations in Euclidean spacetime, they are exponentially unstable when continued back to Lorentzian signature and do not correspond to a sensible theory, i.e. they describe tachyons. The self-accelerated branch is unstable precisely because of these tachyonic modes \cite{greg} and therefore does not admit a meaningful Minkowski space interpretation. We therefore discard this example, but glean from it the diagrammatic roots underlying the running of $G_N$ in the IR for this type of propagator. 

We now turn to another example which manifests novel physics in the IR to gain further traction on our problem-- a class of widely studied (non-local) field theories whose one loop renormalization of gravitational couplings can also be computed. We consider the propagator:
\eq{ncprop}{\frac{1}{\square + \frac{1}{\theta^2\square} - m^2},}
where $\theta$ has dimensions of length$^2$. This propagator can be thought of as arising from loop corrections to a non-commutative scalar field theory in Euclidean spacetime \cite{nd}\cite{min}, or it can be thought of as the free field limit of a translationally invariant, renormalizable non-commutative field theory \cite{adrian}. As before, we are not interested so much in the specific origins of this theory in so much as shall use it as an theoretical laboratory in which test the ideas which we are interested in \footnote{We also note that unlike the  models underlying (\ref{ncprop}), when viewed as a toy example of a {\it free} theory, (\ref{ncprop}) does not have any issues with respect to breaking diffeomorphism or Lorentz invariance.}. It is straightforward to calculate the renormalizations induced by this example (using (\ref{ap5}) in appendix B, and generalizing to the propagator (\ref{ncprop})). We find straightforwardly that 
\eq{nccr}{\lambda(\tau) = \int d^4x~G(x,x;\tau) = \frac{e^{-m^2\tau}}{8\pi^2\theta^2}K_2[2\tau/\theta],}
with $\lambda(\tau)$ again defined as in (\ref{cnomes}) renormalizing the cosmological constant, and with $K_2[x]$ the modified Bessel function of second order. Similarly, Newton's constant is renormalized as
\begin{eqnarray}
\label{ncgn}g(\tau) &=& \frac{e^{-m^2\tau}}{8\pi^2\theta^2} \frac{\tau}{3}\Bigl[K_2[2\tau/\theta] - \frac{\tau}{4\theta}K_3[2\tau/\theta] + \frac{\tau}{4\theta}K_1[2\tau/\theta]\Bigr]\\ \nonumber
&=&  \frac{e^{-m^2\tau}}{8\pi^2\theta^2} \frac{\tau}{6}K_2[2\tau/\theta]
\end{eqnarray}
with $g(\tau)$ defined as in (\ref{gnomes}), and the second line follows from the first via a recursion identity for the modified Bessel functions \cite{AS}. Once more, we find that this model does not exhibit degravitation. This is because the asymptotic form of the modified Bessel functions is exponentially decaying \cite{AS}:
\eq{mbfaf}{\lim_{z\to \infty} K_n(z)\sim \sqrt{\frac{\pi}{2z}}e^{-z},}
and therefore $g(\tau)/\tau$ decays too fast to cause the vanishing of $G_N$ in the deep IR. However by deliberately introducing a tachyonic instability as above, one could again formally induce a degravitation filter, but this time of the massive gravity type (eq. (\ref{rg}) with $\alpha = 1$) \footnote{Again suspending our disbelief, we consider for a moment the case where the momentum space propagator (\ref{ncprop}) can be written in the form: $\Bigl(k - M^2/k\Bigr)^{-2}$, with $M = 1/\theta$, which describes the propagation of massive plane wave modes in Euclidean spacetime-- though as before, merely corresponds to a tachyonic instability ($m^2 = - M^2$). In this case, similarly to the DGP example one can compute the running of $G_N$ and $\Lambda$ to find: $G(\mu) = \frac{G(\mu_r)}{1 + \frac{G(\mu_r)}{6\theta^2}\sqrt{{\pi\theta}}\frac{1}{\mu}}$, and $\Lambda(\mu) = \bar\Lambda(\mu_r) + \frac{1}{2\pi\theta^2}\sqrt{\pi\theta}\mu,$, which would satisfy the requisite properties we require of a degravitating model were it a viable example.}. It seems  neccesary therefore that in order to consistently effect degravitation, one has to consider theories with divergent diagrams that yield the appropriate IR asymptotics for $g(\tau)$ without signalling an instability in the theory at some finite momenta. Moreover this divergence would have to be physical in nature rather than an artifact of any particular scheme. We consider a class of theories which posses the desired features next. 

\section{Degravitation-- a toy model}

We reconsider the general expression for $g(\tau)$, the quantity which determines how $G_N$ is renormalized through (\ref{rungnomes}). Beginning with (\ref{ap5}) in appendix B, denoting $x = k^2$ and performing the angular integrations, we find that
\eq{apx}{g(\tau) = \frac{\tau}{3}\frac{1}{16\pi^2}\int^\infty_0 dx~e^{-\tau u(x)}\Bigl[x u'(x) + u''(x)\frac{x^2}{2} - \frac{\tau}{4}x^2u'(x)^2\Bigr].}
Realizing that the first two terms in the square brackets sum to $(u' x^2/2)'$, and upon integrating by parts, realizing that the last term is minus one half of the sum of the first two terms, permits us to condense all of the above into the expression:
\begin{eqnarray}
\label{apx2}g(\tau) &=& \frac{\tau}{6}\frac{1}{16\pi^2}\int^\infty_0 dx~e^{-\tau u(x)}\frac{d}{dx}\Bigl(\frac{u' x^2}{2}\Bigr)\\
\nonumber &=& \frac{\tau^2}{12}\frac{1}{16\pi^2}\int^\infty_0 dx~e^{-\tau u(x)}u'^2x^2.  
\end{eqnarray}      
We now first perform the $\tau$ integral (which corresponds to a change in the renormalization scheme from before), resulting in:
\eq{tint}{\int^\infty_0 d\tau \frac{g(\tau)}{\tau} = \frac{1}{12}\frac{1}{16\pi^2}\int dx \frac{u'^2x^2}{u^2},}
where now instead we have to regulate the limits of the $x$ integral (recalling $x = k^2$). Considering the example of a canonical scalar field (with $u(x) = x$) yields the standard UV divergence found in (\ref{reng}). Inserting the two different modified propagators considered in the previous section also yields the same UV divergences ($\sim \Lambda^2$ which were absorbed into $G_{ref}$), and give us IR divergences at precisely the same energy scales (at the tachyonic poles), though with different functional forms around this divergence. Thus the inherent scheme dependence of computations in semi-classical gravity rears its head here-- a seemingly inescapable state of affairs (see \cite{percacci} for a discussion). In general, in any enterprise involving computing the running of gravitational couplings such as the one undertaken here, the only strategy we have to proceed with in ensuring the physical sense of our calculations is to repeat them in various different schemes, and at the very least, see if the qualitative nature of the conclusions is scheme independent. We presume that the vanishing of $\Lambda(\mu)G(\mu)/M^2_{pl}$ in the far infra-red (and not at some finite energy scale) is unambiguous enough a feature which one can repeatedly test in different schemes. We bear this important caveat in mind in what follows.

Resuming from (\ref{tint}), in writing our kinetic term as $u(x) = exp[g(x)]$ with $x := k^2$, we see that we require the integrand
\eq{divreq}{\int dx ~[x g'(x)]^2}
to have a divergent lower limit. This would only be possible if $x g'$ diverged as some negative power of $x$ for small $x$. An example of a non-local scalar field theory which accomplishes this is given by the action:
\eq{sfactyes}{S = \int d^4x\sqrt{-g}~\phi[e^{m^2/\square}\square]\phi,}
In spite of its unusual appearances, we motivate this theory shortly \footnote{We note that this theory describes a consistent scalar field which only propagates massless quanta. There is an essential singularity in the propagator at zero momentum, which would only be of concern if it implied a negative residue (and would therefore indicate the presence of ghosts in our system). Although the residue at an essential singularity is indeterminate, one can readily avoid the issue by considering the action in the limit $\epsilon^2\to 0$ of $S = \int d^4x\sqrt{-g}~\phi(e^{\frac{m^2}{\square+\epsilon^2}}~\square]\phi$. As we shall see, such an action can readily be thought of as encoding the non-perturbative resummations of finite renormalization terms.}. Starting with (\ref{tint}), upon regularizing the lower limit of the integral with the IR cut-off $\mu^2$ the generalization of (\ref{gnomex}) becomes:
\eq{gnomex2}{\frac{1}{G(\mu)} = \frac{1}{G(\mu_r)} - \frac{1}{24\pi}\int^{\mu^2_r}_{\mu^2}\frac{u'^2x^2}{u^2}}
where again, the UV divergent contributions have been absorbed into $G_(\mu_r)$ analogously as between (\ref{gnomes}) and (\ref{gnomex}). For $u(x) = xe^{-m^2/x}$, the generalization of (\ref{rungnomes}) becomes:
\eq{rungnomesf}{G(\mu) = \frac{G(\mu_r)}{1 + G(\mu_r)\frac{m^2}{24\pi} \Bigl(\frac{m^2}{\mu^2} - \frac{m^2}{\mu_r^2}\Bigr)}}
which impiles the filter function
\eq{rgFINAL}{G(\square/m^2) = \frac{\bar G(\mu_r)}{1 + \bar G(\mu_r)\frac{m^2}{24\pi} \frac{m^2}{\square}},}
which corresponds to a filter function of the massive gravity type, and where we defined:
\eq{bgdefyo}{\bar G(\mu_r)= \frac{G(\mu_r)}{1 - G(\mu_r)\frac{m^2}{24\pi}\frac{m^2}{\mu_r^2}}.}
We discuss the scale $m$ that appears in the above shortly. The renormalization of the cosmological constant $\Lambda$ is given by:
\eq{lrform}{\Lambda_R = \Lambda_B + \frac{1}{4\pi}\int x dx~ln[u(x)] = \Lambda_B + \frac{1}{4\pi}\int x dx~\Bigl[ ln~x - \frac{m^2}{x}\Bigr].}
The action (\ref{sfactyes}) results in the standard cut-off$^4$ divergence (easily seen from the fact that the kinetic term asymptotes to the standard kinetic term for large momenta), and yields the running in the IR:
\eq{ccrunIRyes}{\Lambda(\mu) = \bar\Lambda(\mu_r) + \frac{m^2}{4\pi}\mu^2,}
with the definition
\eq{predef}{\bar\Lambda(\mu_r):= \Lambda(\mu_r) - \frac{m^2}{4\pi}\mu_r^2.} 
Hence the product $G(\mu)\Lambda(\mu)\to 0$ as $\mu \to 0$, and all other cosmological constant like sources (which will not renormalize $G_N$ in the IR) will be acted upon by the filter function (\ref{rgFINAL}). Therefore this theory appears to offer us a semi-classical realization of degravitation where instead of Einstein gravity, we have its semi-classical incarnation:
\eq{modesemi}{G^{\mu}_\nu = \frac{1}{M^2_{pl}}\frac{1}{1 + \frac{m^2}{24\pi M^2_{pl}} \frac{m^2}{\square}}T^\mu_\nu,}
where we have implicitly taken the scale $\mu_r$ to be such that we measure $1/M^2_{pl}$ as the gravitational coupling at laboratory scale. We note that in order not to affect the behaviour of gravity at sub Hubble scales, we require that the mass scale $m$ in the above be such that
\eq{mconst}{m \leq 10^{-30}M_{pl}\approx 10^{-3} eV,}
whose value we return to discuss shortly. More pressingly, we wish to draw caution against the precise form of (\ref{modesemi}). In the scheme where we compute the running within the proper time formalism (\ref{apx2}) (instead of switching back to momentum space as we did in (\ref{tint})), we find that although degravitation still results, the precise functional form of the running differs. Evaluating the running for $G_N$ that results from (\ref{rungnomes}) and using (\ref{apx2}) with $u(x) = xe^{-m^2/x}$ results in
\eq{ptsrung}{G(\square/\mu_r^2) = \frac{1}{M^2_{pl}\Bigl[1 + \frac{m^2}{24\pi M^2_{pl}} ~ln~[\frac{\mu_r^2}{\square}]\Bigr]},}
and although the integral we have to evaluate to compute the running of the cosmological constant is not possible analytically, we find that $\lambda(t)$ as defined in (\ref{ehren}) numerically evaluates to something that is accurately fit by:
\eq{ltdef}{\lambda(\tau) = \frac{m^4}{16\pi^2}\frac{c}{ln(m^2\tau)},} 
where $c \sim 0.2$ is a numerical co-efficient. This results in the running:
\eq{lbaren}{\Lambda(\mu) = \Lambda(\mu_r) - c \frac{m^4}{8\pi}ln[ln(m^2/\mu^2)],}
from which we see as before that $G_N(\mu)M^2_{pl}\to 0$ and $G_N(\mu)\Lambda(\mu) \to$ as $\mu\to 0$, however the IR asymptotics of the running is markedly slower in this scheme.
Thus, although we have demonstrated by example that renormalization group flow in the IR can indeed result in a filter function for Newton's couping (\ref{gn}) with all the properties required of degravitation, more work remains in order to place this claim on firmer footing. In particular, a full functional renormalization group \cite{cw} analysis might be amenable to the context we are interested in. 
  
We now return to motivating the action (\ref{sfactyes}). We first observe (as is perhaps not widely appreciated) that inverse powers of the d'Alembertian are guaranteed to arise in the IR due to finite terms in the effective action for a scalar field \cite{BM}, where operators of the form
\eq{invops}{\Delta\Gamma \sim \frac{1}{\square}V}
are shown to appear in the IR in interacting massless theories (late times (i.e. large $\tau$) in the so-called proper time formalism used in our heat kernel calculations). Kinetic operators of the form $\square e^{m^2/\square}$ can perhaps then be thought of as modelling a non-perturbative resummation of all such corrections, which we would then subsequently integrate out entirely to result in the running we require. Indeed it is precisely operators of this form that could encode, for example, particle creation effects \cite{BM} as well as vacuum polarization in non-trivial backgrounds-- the consequences of which for dark energy and the cosmological constant problem are discussed in \cite{dynamical}\cite{poly}. Therefore in spite of its unusual appearances, (\ref{sfactyes}) might even be a necessary modification to a QFT interacting with gravity at very long wavelengths. This is our preferred interpretation, though we wish to mention another motivation for (\ref{sfactyes}) which can be inferred from the context of non-commutative field theories. In this context, it is known \cite{min} that the low energy effective action receives non-local loop corrections of the form:
\eq{leeancft}{S = \int d^4x~\phi\Bigl(\square + \frac{1}{\theta^2\square}\Bigr)\phi,}
where $\theta$ has dimensions length squared and parametrizes the degree of spatial graininess \footnote{In addition, if this non-commutativity is restricted to some hidden (i.e. dark) sector which contains the field $\phi$, we bypass having to worry about Lorentz invariance violating effects at low energies.}. Expanding the action (\ref{sfactyes}) at finite momenta results in the expansion:
\eq{leeancft2}{S = \int d^4x~\phi\Bigl(\square + \frac{m^4}{\square} + m^2 + ... \Bigr)\phi,}
from which it is tempting to conclude that (\ref{sfactyes}) might again, be some form of a non-perturbative resummation of loop corrections in a non-commutative field theory \footnote{Furthermore, we note with interest that when viewed in momentum space, the non-local action (\ref{sfactyes}) can even be viewed as the action for a canonical scalar field modified with a suitable IR cut-off function required to compute its functional renormalization group equations \cite{percacci}.}. We also note that the theory described by (\ref{sfactyes}) when considered as the limiting case of a sequence of regularized theories (see footnote under (\ref{sfactyes})), describes a causal theory with a well defined Cauchy problem \cite{neil}\footnote{In \cite{neil} it was shown that it is not the degree of the differential equation defined by a non-local field theory that determines whether or it has a well defined boundary value problem. Rather it is the number of poles, their order, and the residue at the poles of defined by the kernel of the kinetic term in momentum space.}.

Returning now to the scale $m$ in the toy action (\ref{sfactyes}), which we see independent of the precise form of the filter function ((\ref{rgFINAL}) or (\ref{ptsrung})) that results, the requirement that gravity remain unaffected for all sub-Hubble scales requires $m \leq 10^{-3}eV$. We immediately see the naturalness of this requirement depends on the particular physics that gives to (\ref{sfactyes}), however in both contexts described above, it is straightforward to see that this amounts to a technically natural tuning. In the context of having resummed all finite corrections to the effective action of some light scalar degree of freedom (in the dark sector), we see that this scale $m$ is by definition the end result of having summed an infinite tower of quantum corrections. We now offer our concluding thoughts. 

\section{Conclusions}
In this report, we have offered preliminary evidence that degravitation \cite{dgs}\cite{addg} can be rather naturally thought of as an epiphenomenon of semi-classical gravity emergent in the IR. In realizing degravitation as a manifestation of renormalization group flow of gravitational couplings in the IR, its sense as a non-local modification to Einstein gravity, its non-linear completion around arbitrary backgrounds and the question of the modified Bianchi identities that result are readily understood. That it not only solves the bare cosmological constant problem, but also offers us the appropriate dynamics to address the other two cosmological constant problems should serve as strong motivation to further study this mechanism. In particular, that this type of long distance modification to gravity might be forced upon us through non-perturbative resummations of terms that contribute finite renormalizations to the effective action for a canonical scalar field \cite{BM}, is a tempting idea. There is deep conceptual appeal in the idea that the cosmological constant problems might be resolved, rather than aggravated by loop effects in our effective theories.

\section{Acknowledgements}
The author is greatly indebted to Gregory Gabadadze for many useful discussions, hospitality, guidance and collaboration during earlier stages of this work, for which he also wishes to thank Stefan Hofmann. Special thanks to Gia Dvali and Ignatios Antoniadis for discussions, correspondence, and comments on the draft. The author is grateful to Jan Plefka and funds from Humboldt University of Berlin where much of the work that resulted in this report was done. We owe thanks to Robert Brandenberger and John Estes for comments on the draft and many useful discussions, for which we also thank Costas Kounnas and Donovan Young. This work is supported at the Ecole Polytechnique and the Ecole Normale Sup\'erieure with funds from the European ERC Advanced Grant 226371 MassTeV, and from funds from CEFIPRA/IFCPAR. We thank the CCPP at NYU for periodic hospitality over the duration of this project.

\section{Appendix A}
In this appendix, we offer an overview of the mechanisms through which degravitation addresses the various cosmological constant problems. Some of the details concerning what follows can be found in \cite{sp}. We assume a filter function of the form:
\eq{ffgn}{G_N(L^2\square) =  \frac{G_N}{1 + \frac{M^2}{\square}},}
where $G_N = 1/M^2_{pl}$ is the usual laboratory scale 4d Planck mass, and $M$ is a mass scale which defines the IR filter scale $L$ ($M = m \frac{m}{M_{pl}}$ in the examples derived in the previous sections). Clearly given the spectral decomposition of any particular filter function (\ref{specd}), we can easily generalize the results of this section to other functional forms. Equation (\ref{ffgn}) is to be understood in the context of the modified Einstein equations:
\eq{meea}{G^\mu_\nu = 8\pi G_N(L^2\square)T^\mu_\nu.}
To avoid the complexity of tensor structure, we consider the trace of both sides of the above equation in what follows:
\eq{meeat}{R(x) = -8\pi[G_N(L^2\square)T](x)}
Given (\ref{ffgn}) we can recast the above as
\eq{recast}{R + M^2\square^{-1}R = -8\pi G_N T,}
or more concretely:
\eq{recast2}{R(x) = -8\pi G_N T(x) - M^2\int d^4x'\sqrt{-g(x')}G^0(x,x')R(x'),}
where $G^0(x,x')$ is the Green's function for the d'Alembertian operator--
\eq{gfdo}{\square_x G^0(x,x') = \frac{\delta(x,x')}{\sqrt{-g(x')}},}
for which we chose the appropriate causal boundary conditions (so that no sources in the future lightcone of a given spacetime point affects the geometry at that point). With this choice, the support of the integrand in (\ref{gfdo}) extends only over the backward lightcone of the point $x$. One can repeatedly iterate the process above to obtain:
\eq{prep}{R(x) = -8\pi G_N\Bigl[ T(x) -M^2(G^0,T)(x) + M^4(G^0,(G^0,T))(x) -M^6(G^0,(G^0,(G^0,T))(x)\Bigr] + ...,}
where:
\eq{spdef}{(G^0,T)(x):= \int_\blacktriangle d^4x'\sqrt{-g(x')}G^0(x,x')T(x'),}
and the subscript on the integral $\blacktriangle$ emphasizes that the domain of integration is the backward lightcone of the point $x$. Because of this stipulation, the nested integrals in the above are ordered integrals, with each integration variable in the past lightcone of the one proceeding it. We can now proceed to evaluate (\ref{prep}) for various sources.

We commence with a bona fide spacetime zero mode (i.e. the bare cosmological constant). In that case $T = 4\Lambda$, and the nested integrals each evaluate to a constant and can straightforwardly result in:
\eq{nested}{M^{2n}(G^0,(G^0,...,(G^0,T))(x)  = 4\Lambda M^{2n}\Delta^n,}
with
\eq{deldef}{\Delta := \int d^4x'{\sqrt{-g(x')}G^0(x,x')},}
where the domain of integration now stretches infinitely to the past and therefore integrates to (an infinite) constant ($\Delta \to \infty$) \footnote{This feature is essentially due to the fact that interactions mediated by massless fields have divergent scattering cross sections.}. Therefore in resumming (\ref{prep}) we find:
\eq{deldefres}{R = -8\pi G_N \frac{4\Lambda}{1 + M^2\Delta} = 0.}
Thus the bare cosmological constant decouples from geometry. This is perhaps more directly seen from (\ref{ffgn}), where the fact that the d'Alembertian is a self adjoint operator allows us to simply replace the appropriate eigenvalue in place of the operator argument in operator functions. Since in any given spacetime the d'Alembertian always admits constant functions as zero modes (among others), substituting the null eigenvalue in place of $\square$ in (\ref{ffgn}) also demonstrates the desired result.

We now consider a spatially homogeneous source that corresponds to a step function in time in a given FRW slicing. That is, we consider a source of the form:
\eq{sourceeq}{T = -4V \theta(t - t_0).}
This source is to caricature the condensing of a potential dominated epoch at some finite time in the universe's past. Specifically, since we are only considering the trace of the energy-momentum tensor, one can imagine this ansatz to model the onset of inflation from some pre-existing quantum gravity phase, where a radiation dominated epoch makes a transition to a potential dominated epoch at some time $t_0$. We don't tether what follows to any particular scenario, and simply consider (\ref{sourceeq}) as a prototypical source. When substituted into (\ref{prep}) the time ordered nature of the integrals allows us to recast the ordered iteration to obtain the degravitated source as \cite{sp}:
\eq{seq2}{T_{deg}(t) = -4V \theta(t - t_0)\Bigl[1 -M^2\Omega(t) + M^4\frac{\Omega^2(t)}{2!} - M^6\frac{\Omega^3(t)}{3!} + ... \Bigr],}
with $\Omega(t)$ given by
\eq{omdef}{\Omega(t) := \int_\blacktriangle G^0,}
with the domain of integration consisting of the past lightcone of the observer at $t$ up to the nucleation event at $t_0$. Therefore at $t=t_0$, the integrals all vanish given their vanishing domain, after which they steadily increase and asymptote to an infinitely large value. The series (\ref{seq2}) sums readily to give us:
\eq{degsource}{T_{deg}(t) = -4V \theta(t - t_0)e^{-M^2\Omega(t-t_0)}.}
Therefore the curvature sourced by a step function like source (\ref{sourceeq}) is given by \cite{sp}:
\eq{rgen}{R(t) = 8\pi G_N ~4V e^{-M^2\int_\blacktriangle G^0}.}
In its full generality, this equation is a very complicated non-linear integral equation. However we can infer some general features from its form. In particular the argument of the exponential grows as:  
\eq{rgent}{R(t) \sim e^{-M^2(t-t_0)^2},}
where $t-t_0$ is the co-ordinate time that has elapsed since the step function has nucleated \footnote{One might have guessed from dimensional considerations alone that the argument should instead have scaled like $M^2H^2T^4$, where $H$ is some dimensionful parameter characterizing the average expansion of the background. This would follow from the fact that in a background with a curvature set by the scale $H$, we know that $G^0\sim H^2 f(z)$. Although this would have suited the purposes to which we apply our conclusions to {\it even better}, explicit calculations for test sources show that factors of $H$ cancel such that indeed, the decay of a homogeneous source to first approximation is independent of the rate of expansion of the background (see section 5 of \cite{sp}), with decay of the form (\ref{rgent}) resulting. Perhaps easier seen, is that the scaling $\sim e^{-M^2H^2T^4}$ does not yield a sensible $H\to 0$ limit in the sense that the degravitation filter clearly still result in the decay of a test source (\ref{degsource}) in a flat background, whereas the ansatz indicates that this would not be the case.}. This is perhaps easiest seen by considering a test source around a Minkowski background. In that case, $G^0 \sim 1/2\pi^2 r^2$, so that $\int_\blacktriangle G^0 \sim (t-t_0)^2$, resulting in (\ref{rgent}).
From this, we can infer the characteristic timescale
\eq{tscale}{\tau \sim 1/M.}
Imagining now a toy universe where some energy density condensed out of a radiation like epoch according to (\ref{sourceeq}), how long would we have to wait for this energy density to degravitate to the currently observed density that we accord to dark energy? That is, given $V$ of any particular magnitude, how long would we have to wait before this energy density decayed so that $R/M^2_{pl}\sim 10^{-120}$? Evidently, this would be (for initially GUT scale energy densities: $V/M^4_{pl} \sim 10^{-12}$) when the argument of the exponent becomes $\sim 248$.Therefore, 
\eq{exparg}{M^2T^2 \sim 248.}
Given that in the examples considered in the previous sections, $M = m \frac{m}{M_{pl}}$, and given that requiring gravity to remain unaffected up to {\it present} Hubble scale $H_0$ implies (as is easily inferred from (\ref{modesemi}) or (\ref{ptsrung})) that
\eq{hsboun}{\frac{m^2}{M^2_{pl}}\frac{m^2}{H^2_0} \leq 1,}
so that substituting the definition of $M$ in (\ref{exparg}) results in 
\eq{exparg2}{ H_0^2T^2 \geq 248.}
And so,
\eq{exparg3}{T\gtrsim 16 H^{-1}_0.}
Therefore, if we were to degravitate an initial GUT scale energy density all the way down to the currently observed energy density ascribed to dark energy, we would have to wait a period approximately 16 times longer than the current age of the universe. Clearly this particular example does not plausibly account for what dark energy might be. 

However, the non-local nature of the degravitation filter (\ref{ffgn}) results in some very striking features as far as cosmology is concerned. In particular, it implies a memory effect of past energy densities \cite{sp} that might account for the presently observed component that is causing the expansion of our universe to accelerate. To see this, we first note that we can recast the action of (\ref{ffgn}) on some source $\rho$ as:
\eq{rdrewr}{\rho_{degrav}(x) = \rho(x) - M^2\int d^4x'\sqrt{-g(x')}G(x,x')\rho(x'),}
where now $G(x,x')$ is the Greens function for the operator $\square + M^2$, and $\rho_{degrav}(x)$ is the effective source obtained by action of the filter, with $G_N = 1/M^2{pl}$ explicitly factored out as in (\ref{ffgn}). We now consider the effect of a test source, homogeneous in spatial extent but a delta function in time. We work in the test approximation around a fixed de Sitter background with co-ordinate interval:
\eq{crgc}{ds^2 = \frac{1}{H^2\eta^2}(d\eta^2 -dx_idx^i).}
We are interesting in evaluating: 
\eq{thetadeg}{\delta_{degrav}(\eta - \eta_i) = \delta(\eta - \eta_i) - M^2\int d\eta'd^3x' \sqrt{-g(x')}G(x,x')\delta(\eta' - \eta_i).}
Since our source is spatially homogeneous, given the standard expression for the Green's function for a massive field on a de Sitter background (see \cite{sp} for details), we can immediately perform the spatial integrals to yield:
\begin{eqnarray}
\label{spint}\int d\eta'\int d^3x'\sqrt{-g(\eta')}G(x,x')\delta(\eta -\eta_i) &=& \int d\eta'\frac{\theta(\eta - \eta')(-\eta)^{3/2}}{H^2\nu (-\eta')^{5/2}}sinh[\nu ln (\eta/\eta')]\delta(\eta -\eta_i)\\
&=& \nonumber \frac{\theta(\eta - \eta_i)}{2H^2\nu\eta_i}\Bigl[ \frac{(-\eta)^{3/2 - \nu}}{(-\eta_i)^{3/2 - \nu}} - \frac{(-\eta)^{3/2 + \nu}}{(-\eta_i)^{3/2 + \nu}} \Bigr],
\end{eqnarray}
given that conformal time and cosmological time are related via
\eq{ctct}{\eta = -\frac{e^{-Ht}}{H},}
we can rewrite the above as:
\eq{ddegravv}{\delta_{degrav}(\eta - \eta_i) = \theta(\eta-\eta_i)\frac{M^2}{H\nu e^{-Ht_i}}e^{-\frac{3H}{2}(t-t_i)}Sinh[H\nu(t-t_i)],}
with $\nu$ defined as:
\eq{nudef}{\nu = \frac{3}{2}\sqrt{1 - \frac{4M^2}{9H^2}},}
Thus the action of the degravitation on a delta function source is to yield a decaying step function, whose magnitude immediately after the $\eta_i$ is immediately found to be 
\eq{magdelt}{\delta_{degrav}(\eta-\eta_i) \sim \theta(\eta-\eta_i)M^2\Delta(t),}
where $\Delta(t)$ has dimensions of time, but is an otherwise decaying function. Thus the delta function source at time $t_i$ (which can accurately model the trace of the energy momentum tensor of a universe undergoing a burst of inflation from some initial radiation dominated epoch followed by reheating-- provided of course the duration of inflation is far less that $M^{-1}$), under the effects of the degravitation filter, results in an afterglow energy density which would mimic a cosmological constant with a magnitude in Planck units naturally set by the filter scale:
\eq{ecc}{\frac{\Lambda_{eff}}{M^2_{pl}} = \frac{R}{M^2_{pl}} \sim \frac{M^2}{M^2_{pl}} = \frac{m^4}{M^4_{pl}} \lesssim 10^{-120},}
where the latter inequalities follows from the definition $M = m \frac{m}{M_{pl}}$, and the requirement (\ref{mconst}) that the filter function not affect gravity up till horizon scales. In the context where we take the limit $H\to 0$ in the above (so that the background spacetime in which we place this test source tends to Minkowski spacetime), one can interpret this as the generation of an `afterglow' energy density that subsequently sources a de Sitter phase. 

Thus not only is the present scale of dark energy readily explainable within our framework as a memory effect, whose magnitude finds ready, and technically natural explanation as the filter scale $m$, the dynamics also implies that the co-incidence problem is greatly alleviated. The reason for this is that even though vacuum energy densities do not scale with the expanding spacetime, they decay over time as a result of degravitation and consequently, the coincindence problem (which obtains its severity from the fact that vacuum energy densities do not redshift) is greatly ameliorated \cite{sp}. Thus we hope to have demonstrated that in making the simple assumption that long wavelength sources decouple from sourcing spacetime curvature, all three cosmological constant problems are readily, and naturally addressed.

\section{Appendix B}
In what follows, we generalize the derivation of the heat kernel (\ref{cshk}) adapted to propagators with modified kinetic terms. For concreteness, we begin with the example of a DGP like propagator (\ref{modprop}) but immediately generalize our results to more general functions of $\square$. Our treatment borrows from that reviewed in \cite{pb}. Working in the expanded basis $|x,\tau\rangle$, where we augment the usual 4-d position eigenkets $|x\rangle$ with the basis for $\tau$ space $|\tau\rangle$, we see that in order to compute the heat kernel (\ref{hk}), we need to evaluate:
\eq{csgfi}{G(x,y;t) = \langle x,t| \frac{1}{\partial_t - [\square - M \sqrt{-\square}] + m^2} |y,0\rangle.}
We realize that we can expand the operator in the denominator in a curvature expansion around a locally flat coordinatization of our manifold as:
\begin{eqnarray}
\label{cegf}\partial_t - [\square - M\sqrt{-\square}] + m^2 &=& \partial_t -[\square_0 - M \sqrt{-\square_0}] + m^2 ~~~~~~~~~~~\}A\\
\nonumber
&-& \Bigl(1 + \frac{M}{2\sqrt{-\square_0}} \Bigr)[a^{\mu\nu}\partial_\mu\partial_\nu + b^\mu\partial_\mu] ~~~~\}B   
\end{eqnarray}
where $\square_0 = \partial_\mu\partial^\mu$ is the usual flat space d'Alembertian, and where
\eq{dbd}{\delta\square:= [a^{\mu\nu}\partial_\mu\partial_\nu + b^\mu\partial_\mu].}
The inverse of the operator grouping $A$ is simply the Green's functions in flat coordinates that we compute presently. By (\ref{pde}), we see that we first have to find the appropriate Green's function for the diffusion equation implied by the modified propagator in flat coordinates. Writing the heat kernel as $G(x,x';\tau)= \bar G(x,x';\tau)\theta(\tau)$, one has the following boundary value problem to solve:
\begin{eqnarray}
\label{mpde}
\bar G(x,x';0) &=& \delta^4(x-y)\\
\nonumber
\dot{\bar G}(x,x';\tau) &=& \Bigl(\square_x - M\sqrt{-\square}_x - m^2\Bigr)\bar G(x,x';\tau).
\end{eqnarray}
By translational invariance, we Fourier decompose $\bar G$ and obtain the following solution which satisfies the above with the appropriate boundary value:
\begin{eqnarray}
\label{solnfs}
G(x,y;\tau) &=& \frac{e^{-m^2\tau}}{(2\pi)^4}\int d^4k e^{-ik\cdot(x-y)}e^{-\tau[k^2 + M k]}\\
\nonumber
&=& \frac{e^{-m^2\tau}}{4\pi^2}\int dk \frac{J_1(kr)}{kr}e^{-\tau[k^2 + M k]}k^3,
\end{eqnarray}
Where $J_1$ is the first order Bessel function of the first kind. In the absence of the term proportional to $M$, one straightforwardly recovers (\ref{gffp}). When $y=x$, we can easily evaluate (\ref{solnfs}) to yield
\eq{gfa}{G(x,x;\tau) = \frac{e^{-m^2\tau}}{16\pi^2} \Bigl[ \frac{1}{\tau^2} + \frac{M^2}{4\tau} \mp M \frac{e^{\frac{M^2\tau}{4}}\sqrt{\pi\tau}(6 + M^2\tau) [1 \mp Erf (M \sqrt\tau /2)]} {8\tau^2}\Bigr],}
where the different signs correspond to the choices $\square \pm M\sqrt{-\square}$ for the kinetic term. We note that implicit in (\ref{pde}) is the fact that we are working in units where the diffusion co-efficient is taken to be unity, thus giving $\tau$ dimensions of inverse mass squared. The function $\lambda(\tau)$ (which determines the renormalization of $\Lambda$ through (\ref{cnomes})) is given by (\ref{gfa}). 

Proceeding, we realize that in a locally flat (Riemannian normal) co-ordinatization around the point $x$, one can expand the metric as
\eq{metexp}{g_{\mu\lambda}(x') = \eta_{\mu\lambda} - \frac{1}{3}R_{\mu\nu\lambda\beta}(x)y^\nu y^\beta,}
where $y^\mu = x'^\mu - x^\mu$-- allowing us to compute the operator coefficients in operator grouping $B$ as:
\begin{eqnarray}
\label{a}
a^{\mu\nu} &=& \frac{1}{3}R^{\mu~\nu}_{~\rho~\lambda}y^\rho y^\lambda,\\
\label{b}
b^{\mu}&=& -\frac{2}{3}R^\mu_\lambda y^\lambda .
\end{eqnarray}    
Performing the expansion:
\eq{csgfii}{G(x,y;t) = \langle x,t| \frac{1}{A - B}|y,0\rangle = \langle x,t| \frac{1}{A} + \frac{1}{A}B\frac{1}{A} |y,0\rangle,}
which is equivalent to a curvature expansion, we can evaluate the curved space modified heat kernel as
\eq{csmhk}{G(x,y;t) = \langle x,t| A^{-1}|y,0\rangle + \int^t_0\int d^4z\langle x,t| A^{-1}|z,\tau\rangle\langle z,\tau| BA^{-1}|y,0\rangle.}
Denoting the flat space modified Green's function (\ref{solnfs})(\ref{gfa}) as $\mathcal G(x,x';t)$, we rewrite the above as
\begin{eqnarray}
\label{csmhkr}G(x,y;t) = \mathcal G(x,y;t) + \int^t_0 d\tau\int d^4z \mathcal G(x,z;t-\tau)\Bigl[a^{\mu\nu}(z)\frac{\partial}{\partial z^\mu}\frac{\partial}{\partial z^\nu} + b^\mu(z)\frac{\partial}{\partial z^\mu} \Bigr]\Bigl(1 + \frac{M}{2\sqrt{-\square_z}} \Bigr)\mathcal G(z,y;\tau),
\end{eqnarray}
where first term in the above for $x = x'$ is given by (\ref{gfa}). We next consider the second term in the above. We begin by inserting (\ref{a})(\ref{b}), to obtain:
\eq{ap1}{\int^t_0 d\tau\int d^4z \mathcal G(x,z;t-\tau)\Bigl[\frac{1}{3}R^{\mu~\nu}_{~\rho~\lambda}z^\rho z^\lambda\frac{\partial}{\partial z^\mu}\frac{\partial}{\partial z^\nu} -\frac{2}{3}R^\mu_\lambda z^\lambda\frac{\partial}{\partial z^\mu} \Bigr]\Bigl(1 + \frac{M}{2\sqrt{-\square_z}} \Bigr)\mathcal G(z,y;\tau).}
We now substitute the Fourier expansions (\ref{solnfs}) to obtain:
\begin{eqnarray}
\label{ap2}
\int^t_0 d\tau\int d^4z \frac{e^{-m^2t}}{(2\pi)^8}&&\int d^4k \int d^4p e^{-ik\cdot(x-z)}e^{-(t-\tau)[k^2 + M k]}e^{-\tau[p^2 + M p]}\\
&\times & \Bigl[\frac{1}{3}R^{\mu~\nu}_{~\rho~\lambda}z^\rho z^\lambda\frac{\partial}{\partial z^\mu}\frac{\partial}{\partial z^\nu} -\frac{2}{3}R^\mu_\lambda z^\lambda\frac{\partial}{\partial z^\mu} \Bigr]\Bigl(1 + \frac{M}{2\sqrt{-\square_z}} \Bigr)e^{-ip\cdot(z-y)}
\end{eqnarray}
In order to proceed in a more orderly manner, we introduce the function 
\eq{ul}{u(s):= s + M \sqrt{s},}
so that we can express the above as:
\begin{eqnarray}
\label{ap2.1}\nonumber g(\tau) =
\int^t_0 d\tau\int d^4z \frac{e^{-m^2t}}{(2\pi)^8}&&\int d^4k \int d^4p e^{-ik\cdot(x-z)}e^{-(t-\tau)u(k^2)}e^{-\tau u(p^2)}\\
&\times & \Bigl[\frac{1}{3}R^{\mu~\nu}_{~\rho~\lambda}z^\rho z^\lambda\frac{\partial}{\partial z^\mu}\frac{\partial}{\partial z^\nu} -\frac{2}{3}R^\mu_\lambda z^\lambda\frac{\partial}{\partial z^\mu} \Bigr]u'(p^2)e^{-ip\cdot(z-y),}
\end{eqnarray}
where the prime refers to the derivative of (\ref{ul}) with respect to its argument, and we identify $g(\tau)$ as the co-efficient of the spacetime integral of the Ricci scalar as defined in (\ref{ehren}). In this form, we can straightforwardly consider modified propagators of more general functional forms. We now act on the exponent on far right with the $z$ derivatives, and write $z_\rho$ as $i\partial/\partial p^\rho$, integrate by parts over the $p$ variables, and perform the $z$ integral to obtain a delta function over $p$, resulting in:
\begin{eqnarray}
\label{ap3} g(\tau) =
\int^t_0 d\tau \frac{e^{-m^2t}}{(2\pi)^4}\int d^4k e^{-ik\cdot(x-y)}e^{-(t-\tau)u(k^2)}&\Bigl[& \frac{\partial}{\partial k^\lambda} \Bigl(\frac{2}{3}u'(k^2) R^\mu_\lambda k_\mu  e^{-\tau u(k^2)} \Bigr)\\ &+&\nonumber \frac{1}{3}\frac{\partial}{\partial k^\rho}\frac{\partial}{\partial k^\lambda}\Bigl(R^{\mu~\nu}_{~\rho~\lambda} k_\mu k_\nu e^{-\tau u(k^2)}u'(k^2)\Bigr)\Bigr].
\end{eqnarray}
In the second term in the above, we note by the symmetries of the Riemann tensor that the $k^\lambda$ derivative can only act on $k^\mu$ (else there will be a term symmetric in $k^\nu k^\lambda$ contracted with $R^{\mu}_{\rho\nu\lambda}$), resulting in: 
\begin{eqnarray}
\label{ap3.1} g(\tau) = 
\int^t_0 d\tau \frac{e^{-m^2t}}{(2\pi)^4}\int d^4k e^{-ik\cdot(x-y)}e^{-(t-\tau)u(k^2)}\frac{\partial}{\partial k^\lambda} \Bigl(\frac{u'(k^2)}{3} R^\mu_\lambda k_\mu  e^{-\tau u(k^2)} \Bigr),
\end{eqnarray}
which evaluates as:
\eq{ap5}{ g(\tau) = \frac{e^{-m^2t}}{(2\pi)^4}\int d^4k~ e^{-t u(k^2) }\Bigl[\frac{u'(k^2)}{3} Rt +\frac{2}{3} t u''(k^2) R^\mu_\lambda k_\mu k^\lambda - \frac{t^2}{3}u'^2(k^2) R^\mu_\lambda k_\mu k^\lambda\Bigr],}
where we have performed the trivial $\tau$ integrals and set $x = y$. This is a general expression valid for any form of the kinetic operator $u(k^2)$. We now realize that we can work in a basis in $k$ space where the Ricci tensor is diagonal. Furthermore, we note that one can decompose the Ricci tensor in this basis as:
\eq{riccidec}{R^\mu_\nu = \frac{R}{4}\delta^\mu_\nu + c^\mu_\nu,}
where $c^\mu_\nu$ is diagonal and traceless, and does not renormalize the Einstein Hilbert term (in addition to vanishing upon integration in isotropic though not necessarily homogeneous spacetimes). Substituting $R^\mu_\nu = \delta^\mu_\nu R/4$ into the above, we then arrive at the expression for a DGP-like propagator:
\eq{finexp}{g(\tau) = \frac{e^{-m^2 t}}{8\pi^2}\int^\infty_0 k^3 dk e^{-t[k^2 + Mk]}\Bigl[\frac{1}{3}\Bigl(1 + \frac{M}{2k}\Bigr) Rt - \frac{M}{6\cdot 4k} Rt - \frac{1}{12}\Bigl(k + \frac{M}{2}\Bigr)^2Rt^2\Bigr],}
from which, it is a straightforward process to compute the term in the heat kernel proportional to the spacetime integral of the Ricci scalar as:
\eq{finexp2}{g(\tau) = \frac{e^{-m^2t}R}{16\pi^2}\Bigr[\frac{1}{6t} + \frac{M^2}{96}  - \frac{M}{32 t}[1 - Erf(M\sqrt t/2)\sqrt{\pi t} e^{M^2t/4}]\Bigl(\frac{M^2t}{6} + 3\Bigr)\Bigr].}
Furthermore, we note that the renormalizations induced by the propagator $\square +M\sqrt{-\square}$ (i.e. with the other branch of the square root) can be similarly calculated, and found to result in the same expression above but with the naive substitution $M\to -M$:
\eq{finexp3}{g(\tau) = \frac{e^{-m^2t}R}{16\pi^2}\Bigr[\frac{1}{6t} + \frac{M^2}{96}  + \frac{M}{32 t}[1 + Erf(M\sqrt t/2)\sqrt{\pi t} e^{M^2t/4}]\Bigl(\frac{M^2t}{6} + 3\Bigr)\Bigr].}

\end{document}